\documentclass{optica-article}

\journal{opticajournal} 

\articletype{Research Article}
\usepackage{amsmath} 
\usepackage{lineno}
\usepackage{bm}
\usepackage{threeparttablex}

\begin{document}

\title{Optical turbulence profiling at the Table Mountain Facility with the \textit{Laser Communication Relay Demonstration} GEO downlink}

\author{Marcus Birch,\authormark{1,2,*} 
Sabino Piazzolla,\authormark{2} 
Preston Hooser,\authormark{2}
Francis Bennet,\authormark{1}
Tony Travouillon,\authormark{1}
William Buehlman,\authormark{2}}

\address{\authormark{1}Advanced Instrumentation Technology Centre, Research School of Astronomy and Astrophysics, Australian National University, Canberra, ACT 2611, Australia\\
\authormark{2}Jet Propulsion Laboratory, California Institute of Technology, Pasadena, CA, USA 91109}

\email{\authormark{*}marcus.birch@anu.edu.au} 



\begin{abstract*}

We report the first measurement of the atmospheric optical turbulence profile using the transmitted beam from a satellite laser communication terminal. A Ring Image Next Generation Scintillation Sensor (RINGSS) instrument for turbulence profiling, as described in Tokovinin (\textit{MNRAS}, 502.1, 2021, 747-808), was deployed at the NASA/Jet Propulsion Laboratory's Table Mountain Facility (TMF) in California \cite{tokovinin2021measurement}. The optical turbulence profile was measured with the downlink optical beam from the \textit{Laser Communication Relay Demonstration} (LCRD) Geostationary satellite. LCRD conducts links with the Optical Communication Telescope Laboratory ground station and the RINGSS instrument was co-located at TMF to conduct measurements. Turbulence profiles were measured at day and night and atmospheric coherence lengths were compared with other turbulence monitors such as a solar scintillometer and Polaris motion monitor. RINGSS sensitivity to boundary layer turbulence, a feature not provided by many profilers, is also shown to agree with a boundary layer scintillometer at TMF ($R=0.85$). Diurnal evolution of optical turbulence and measured profiles are presented. The correlation of RINGSS with other turbulence monitors ($R=0.75-0.86$) demonstrates the concept of free-space optical communications turbulence profiling, which could be adopted as a way to support optical ground stations in a future Geostationary feeder link network. These results also provide further evidence that RINGSS, a relatively new instrument concept, correlates well with other instruments in daytime and nighttime turbulence.
\end{abstract*}

\section{Introduction}

The prescence of optical turbulence degrades free-space optical communication (FSOC) links, particularly for high data rates and coherent links where technologies like adaptive optics (AO) are often employed to compensate \cite{piazzolla2023ground,schieler2023orbit}. Measurements from monitoring atmospheric turbulence have many applications to FSOC, such as supporting the operation and design of AO systems, forecasting of turbulence conditions, or refining link budgets \cite{osborn2018atmospheric,basu2020mesoscale}. Improving AO systems will lead to more reliable links by stabilising and boosting received power. High data rate modems require the coupling of the downlink signal into a single mode fiber and are therefore especially sensitive to the received point spread function (PSF). In these configurations, the use of AO brings the downlink PSF closer to the diffraction limit and improves the coupling efficiency \cite{piazzolla2023ground,schieler2023orbit}. Turbulence forecasting can also be used with a network of optical ground stations (OGS) to help plan space-to-ground links based on future conditions such as received power. Although more complex, resolving the vertical profile of atmospheric turbulence is a superior tool to measuring integrated turbulence for all these applications. The turbulence profile contains more information which can be used to derive unique parameters such as the isoplanatic angle and coherence time, both of which are valuable for designing uplink and adaptive optics systems.

Geostationary orbit (GEO) feeder links have been proposed as way to implement global FSOC networks and numerous countries are progressing this capability \cite{hauschildt2019hydron,kotake2022status,kolev2022preparation}. The \textit{Laser Communications Relay Demonstration} (LCRD) is a payload that was launched to GEO in 2021 onboard the STPSat-6 satellite to demonstrate a FSOC relay, conducting simultaneous bi-directional links with the Optical Communications Telescope Laboratory (OCTL) at Table Mountain Facility (TMF) and another OGS at Haleakala, Hawaii \cite{israel2023early}.

We demonstrate the concept of FSOC turbulence profiling by using the downlink signal from LCRD near $1550$~nm as opposed to celestial objects typically used for turbulence monitors. The narrow linewidth from the satellite downlink allows narrowband filtering for 24~hr measurements and high signal-to-noise down to very shallow solar-separation angles while the stability of GEO allows for long observation periods with minimal pointing and tracking.  Furthermore, FSOC turbulence profiling inherently observes through a slant path parallel to the OGS conducting the link, which is beneficial when interpreting these results for OGS operation. Real-time information about the atmospheric channel could be used for turbulence forecasting and to optimise AO performance. A space terminal could also adapt bit rate depending on the received power from ground terminal forecasts.

We have deployed a Ring Image Next Generation Scintillation Sensor (RINGSS) at TMF to conduct the turbulence profiling with the LCRD downlink beam. RINGSS, first outlined in \cite{tokovinin2021measurement} as an evolution of the well established MASS profiler, operates by capturing defocused ring images at high speed. Scintillation is imaged as azimuthal variation around the defocused ring image and the process of estimating a turbulence profile from this is explained in \autoref{sec:background}.  Simpler methods exist for measuring integrated turbulence, such as solar scintillometers, DIMMs, or motion monitors (introduced in \autoref{sec:accuracy}), but do not provide the additional capability provided from resolving the vertical profile of turbulence. RINGSS is novel as a profiler that can be implemented with a telescope as small as 12~cm using commercial off-the-shelf components, unlike the more well-known MASS which requires many custom components, or SCIDAR which requires a $\gtrapprox1$~m telescope \cite{avila1997whole,kornilov2003mass}. A number of other instruments, namely the SHIMM and FASS, have also been developed recently and show promise for turbulence profiling with reduced complexity \cite{griffiths2024comparison,guesalaga2021fass}.

\section{RINGSS at Table Mountain Facility}

\begin{figure}
  \centering
  \includegraphics[width=\linewidth]{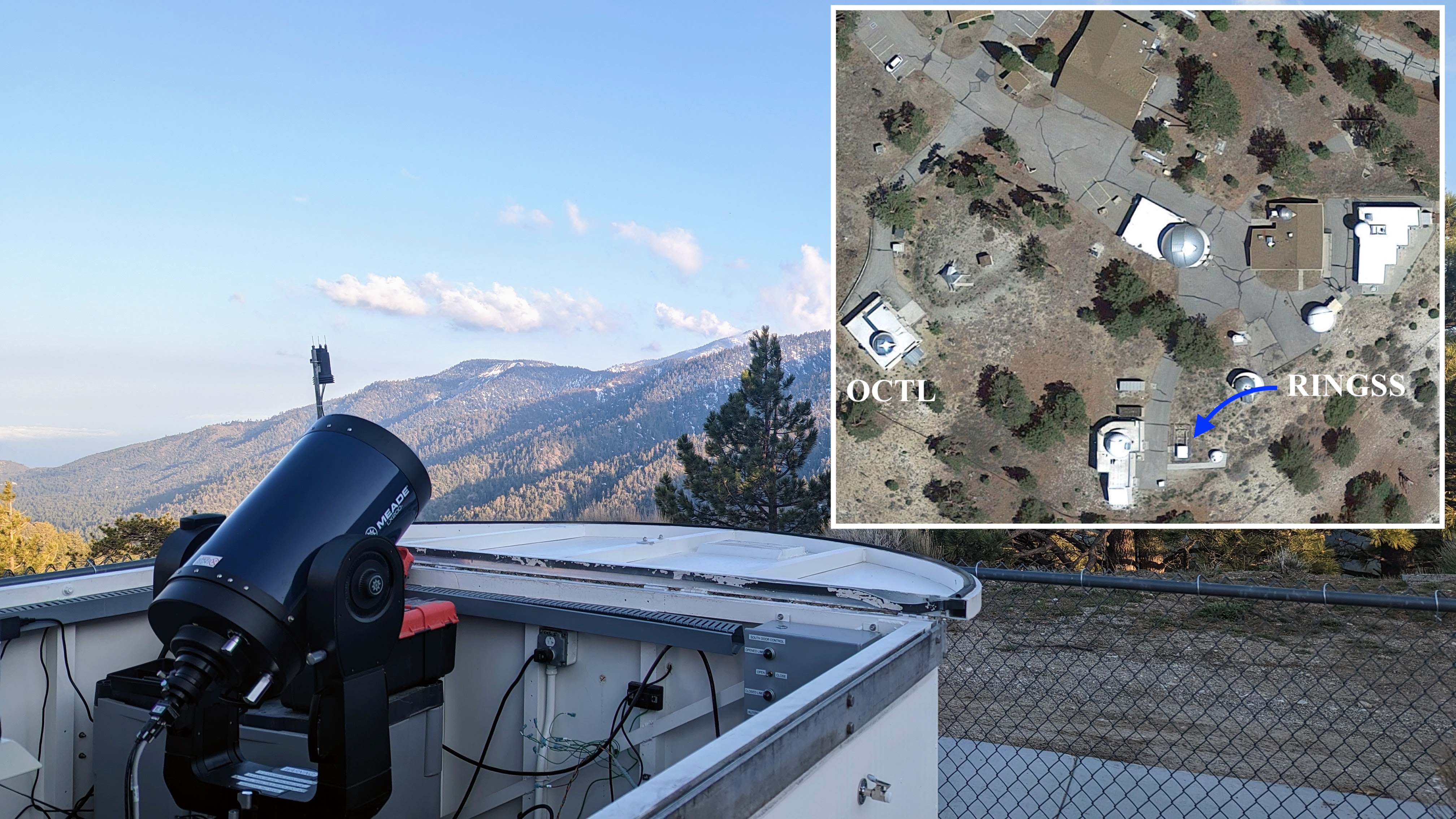}
  \caption{RINGSS instrument with sliding-roof enclosure at Table Mountain Facility, California. Inset map shows proximity to Optical Communications Telescope Laboratory (OCTL).}
  \label{fig:ringss_image}
\end{figure}

The RINGSS instrument implemented for this experiment uses a 25~cm Schmidt-Cassegrain telescope and a 320x256 InGaAs array with reimaging optics to induce the required image specifications. The system produces a ring image at the InGaAs array obtained by using a high secondary obstruction (50\% by diameter) which is achieved with a 3D printed aperture mask.

A narrow linewidth filter at $1550$~nm is placed in the lens barrel before the reimaging optics, ensuring high signal strength at all times of day and even with shallow solar separation. The bright downlink beam from LCRD allows 200~\textmu s exposure times, though the frame rate is limited to 700~Hz. This very short exposure time reduces the blurring of turbulent motion to improve accuracy, particularly in the presence strong turbulence. We house the instrument in a sliding-roof enclosure at TMF close to OCTL, shown in \autoref{fig:ringss_image}. This proximity ensures both systems are observing through a similar atmospheric channel. As LCRD is in a equatorial geostationary orbit, both RINGSS and OCTL statically point at approximately 50$^\circ$ in elevation during links with minimal tracking and pointing required. However, the optics of this system induces strong coma that requires tight centering within $\pm10$~arcseconds of the optical axis.

\begin{figure}
  \centering
  \includegraphics[width=0.5\linewidth]{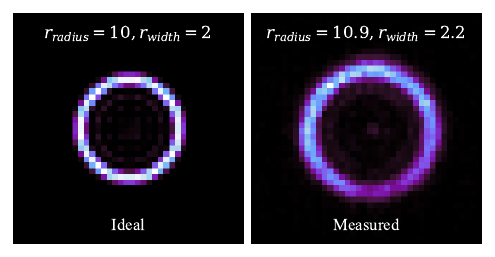}
  \caption{Comparison between the theoretical and measured ring image produced from the defocused LCRD signal using RINGSS instrument. Measured data here is in pixel units and only a small time-averaged sample.}
  \label{fig:rings_comp}
\end{figure}

Although the LCRD downlink is bright enough to easily observe the ring with a smaller aperture, constraints of sizing the ring image with small format and large pixel size (12.5~\textmu m) InGaAs arrays makes a 25cm telescope worthwhile. The RINGSS reimaging optics resize the pixel scale to achieve an optimal conjugation height as described in \cite{tokovinin2021measurement}. The optics are also designed to achieve the $1:10$ ratio of spherical aberration to defocus necessary for a conic wavefront, i.e. producing a close to ``diffraction-limited'' ring width of 2 pixels (\autoref{fig:rings_comp}). Ring width ($R_\text{width}$) and radius ($R_\text{radius}$) are critical parameters for the instrument, the latter is kept to $10\pm1$ pixels while $R_\text{width}$ is generally maintained as $<3$ pixels. \autoref{fig:rings_comp} shows an idealised ring image from our specifications compared with a sample of data from LCRD. Image quality parameters, e.g. $R_\text{width}$, $R_\text{radius}$, and coma, are used in the software as inputs for auto-focusing, tracking, or rejecting erroneous data.

\subsection{RINGSS background}
\label{sec:background}

RINGSS deployed at TMF captures defocused ring images of the LCRD downlink at high speed ($700$~Hz), measuring scintillation as azimuthal variations of intensity around the ring over time. Scintillation around the ring image is decomposed into angular frequencies which when combined with theoretical relations for how different frequencies propagate, allows for the optical turbulence profile to be estimated. The frequency spectrum of scintillation relates to the perturbed wavefront having travelled through the atmospheric column because a defocused image approximates the pupil. A brief explanation of the operational principle is provided below but see \cite{tokovinin2021measurement} for a comprehensive explanation. The angular power spectrum (APS), as measured from scintillation around the ring, is retrieved with annular masks corresponding to discrete angular frequencies. For a given angular frequency, $m$, masks applied to the high speed series of images produce a term, $\langle\vert a_m \vert^2\rangle$, to recover the power of that frequency \cite{tokovinin2021measurement}. The value $\langle\vert a_m \vert^2\rangle$ can be thought of as the azimuthal temporal variance through a frequency filter, hence the set of $\langle\vert a_m \vert^2\rangle$ over a range of $m$ represents the APS. $200$~\textmu s exposures do not warrant a finite-exposure time correction, as used by \cite{tokovinin2021measurement} for 1~ms exposures, $\langle\vert a_m \vert^2\rangle$ values are adjusted for a detector read noise of $30$~$e^-$. The APS is measured in real-time every few seconds from sets of 2000 images.

\begin{figure}[h]
  \centering
  \includegraphics[width=0.7\linewidth]{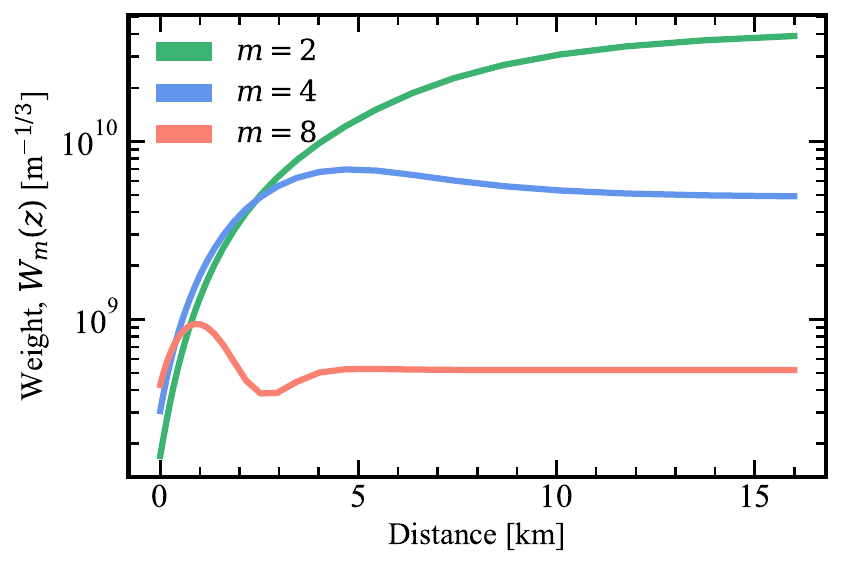}
  \caption{Weighting functions, $W_m(z)$, from $0$ to $16$~km for $m=2,4,8$.}
  \label{fig:WFs}
\end{figure}

With the APS measured, RINGSS then relies on a principle similar to MASS and other scintillation-based turbulence profilers, weighting functions can be formulated to describe how spatial frequencies propagate and instrument response to those optical aberrations. Optical turbulence at any altitude is assumed to have a Kolmogorov spectrum with a strength given by the coefficient of the refractive index structure function, $C_n^2$, in units $m^{-2/3}$. The general principle for weighting functions is that higher frequencies decrease with distance faster than lower frequencies, such that the shape of the APS can indicate the height of turbulence \cite{kornilov2003mass,kornilov2007combined,tokovinin2007accurate,tokovinin2008fade,tokovinin2021measurement}. The process of computing weighting functions is best explained by Tokovinin in \cite{tokovinin2021measurement}, however an important consideration here is that the LCRD downlink is monochromatic, simplifying the process by ignoring polychromatic corrections implemented for prior instruments \cite{tokovinin2003polychromatic,tokovinin2007accurate}. Weighting functions are also dependent on conjugation height, itself a function of defocus or the ring radius, e.g. $R_\text{radius}=10\pm1$ pixels corresponds to conjugation heights of $\approx-780\pm100$~m, so drifts in defocus can lead to substantial error particularly for turbulence near the surface. We minimise this by building a look-up table of $W_m(z)$ for different conjugation heights, which is matched to $R_\text{radius}$ measured in real-time. \autoref{fig:WFs} shows weighting functions for $m=2, 4$ and $8$ at the nominal defocus ($R_\text{radius}=10$ pixels).

We use these weighting functions known \textit{a priori} and the APS measured in real-time, to then consider the atmospheric turbulence profile as $N$ discrete columns of turbulence with strength $J_i=\int_{z_i}^{z_{i+1}} C_n^2(h)dh$ in m$^{1/3}$ that can be solved as a linear system. A set of altitudes $\{z_0,...,z_i,...,z_{N-1}\}$ with $N=8$ from 0 to 16~km is used in keeping with the vertical resolution of the instrument put forth in \cite{tokovinin2021measurement} optimised for instrument response. \autoref{eq:linear_system} shows the linear system which is solved using non-negative least squares to recover the vector of $J_i\forall z_i$, i.e. the optical turbulence profile. This discrete layered approach is also used in practice for numerical propagation methods for simulating links through the atmosphere \cite{townson2019aotools,farley2023fast}. For a given $m$, \autoref{eq:linear_system} can be written as \autoref{eq:main_eq} with rows of the two-dimensional $W_m(z_i)$ matrix becoming functions as shown in \autoref{fig:WFs}. The maximum angular frequency, $M$, is constrained by the Nyquist sampling criteria of two pixel resolution in azimuth, which is $\approx30$ for $R_\text{radius}=10$ pixels, though we set $M=20$ due to noise near this upper limit \cite{tokovinin2021measurement}.

\begin{equation}
    \begin{bmatrix}
        \langle\vert a_{0}\vert^2\rangle \\
        \langle\vert a_{1}\vert^2\rangle \\
        \vdots \\
        \langle\vert a_{m}\vert^2\rangle \\
        \vdots \\
        \langle\vert a_{M}\vert^2\rangle
    \end{bmatrix}
    =
    \begin{bmatrix}
        W_{0}(z_0) & W_{0}(z_1) & \cdots & W_{0}(z_i) & \cdots & W_{0}(z_{N-1}) \\
        W_{1}(z_0) & W_{1}(z_1) & \cdots & W_{1}(z_i) & \cdots & W_{1}(z_{N-1}) \\
        \vdots & \vdots & \ddots & \vdots & \ddots & \vdots\\
        W_{m}(z_0) & W_{m}(z_1) & \cdots & W_{m}(z_i) & \cdots & W_{m}(z_{N-1})\\
        \vdots & \vdots & \ddots & \vdots & \ddots & \vdots\\
        W_{M}(z_0) & W_{M}(z_1) & \cdots & W_{M}(z_i) & \cdots & W_{M}(z_{N-1})
    \end{bmatrix}
    \begin{bmatrix}
        J_{z_0} \\
        J_{z_1} \\
        \vdots \\
        J_{z_i} \\
        \vdots \\
        J_{z_{N-1}}
    \end{bmatrix}
    \label{eq:linear_system}
\end{equation}

\begin{equation}
   \langle\vert a_m\vert^2\rangle =\sum_{i=0}^{N-1}J_iW_m(z_i)
   \label{eq:main_eq}
\end{equation}

Once we derive the set of $J_i$, we compute a number of integrated parameters from the turbulence profile. The atmospheric coherence length or Fried parameter ($r_0$) describes the general turbulence conditions and is used for comparison with other turbulence monitors. The method of computing $r_{0}$ from $J_i$ is given in \autoref{eq:r0scint} and referred to as $r_{0,\text{scint}}$ to differentiate from the separate method of estimating $r_0$ described in \autoref{sec:radial}. The value $\cos{\left(\zeta\right)}\sum_{i=0}^{N-1} J_i$ in \autoref{eq:r0scint} represents the airmass-corrected integral of the turbulence profile with zenith angle, $\zeta$, which is approximately $40^\circ$ for TMF observing LCRD in Geostationary orbit. The $\cos{(\zeta)}$ term takes the turbulence measured through the slant path and scales it to represent turbulence at zenith. All $r_0$ values throughout this paper are reported for $\lambda=500$~nm and airmass-corrected to zenith in keeping with the convention for generalising turbulence conditions. Other parameters such as the atmospheric coherence time and isoplanatic angle can also be recovered from the turbulence profile and used for additional purposes such as informing system design, e.g. AO speed requirements or point-ahead compensation as discussed in \autoref{sec:prof_results}. 

\begin{equation}
    r_{0,\text{scint}}=\left(16.7\lambda^{-2}\cos{\left(\zeta\right)}\sum_{i=0}^{N-1} J_i\right)^{-3/5}
    \label{eq:r0scint}
\end{equation}

\subsection{Radial motion turbulence measurement}
\label{sec:radial}

Radial motion around the ring can be used to measure the integrated turbulence strength in a manner similar to a Differential Image Motion Monitor (DIMM), in addition to the primary method discussed previously \cite{tokovinin2002differential}. This technique involves splitting the ring into eight radial sectors of equal azimuth and observing the variance of the ring radii from sets of opposing sectors, i.e. 4 directions of motion.  Radial motion is differential in the sense it is relative to the ring's center, making it independent of wind shake and other phenomena but sensitive to ring width calibration to prevent motion blurring. Variance of these four opposing sector sets is analogous to the two perpendicular directions used for measuring differential motion in a DIMM. Masks corresponding to each sector are used to compute the temporal variance, $\sigma_{2r}^2$, similar to how the APS is computed from azimuthal motion. The four sets of opposite sectors are averaged to estimate $\sigma_{2r}^2$. A normalisation constant for sector weights, $C_r$, is then used as in \autoref{eq:sector seeing} to compute $r_{0,\text{sector}}$, where $D$ is the telescope diameter. See \cite{tokovinin2021measurement} for further information on $\sigma_{2r}^2$ and $C_r$, where the latter is similar to the normalised weighting coefficient used for DIMMs. $r_{0,\text{sector}}$ is reported for $\lambda=500$~nm and air-mass corrected to zenith in the same way as $r_{0,\text{scint}}$, although $\sigma_{2r}^2$ and $C_r$ are computed with the $1550$~nm wavelength of operation.

\begin{equation}
    r_{0,\text{sector}}=0.98\lambda^{4/5}\left(\frac{\sigma_{2r}^2}{4C_r}\right)^{3/5}D^{1/5}
    \label{eq:sector seeing}
\end{equation}

\section{Calibration}
\label{sec:calib}

A preliminary dataset with RINGSS at TMF of approximately 5 days throughout May and June of 2023 is used in this paper over a number of LCRD links that were conducted sporadically at daytime and nighttime. These observations correspond to approximately twelve hours of turbulence profile data. In this section we explain how a number of parameters such as $R_\text{width}$, $R_\text{radius}$, and coma can be used to assess data reliability or tune operation. While inputs for focus and pointing control loops are important, maintaining data integrity throughout the processing pipeline is also critical as sudden drops in quality can be common, e.g. from very high winds near the telescope.

Once data cubes are captured with RINGSS and the APS is measured, images outside the specification of $9<R_\text{radius}<11$ and $R_\text{width}<3.5$, are flagged as low quality and ignored when computing profiles for real-time display of the turbulent conditions (but are saved nonetheless for analysis of instrument performance such as presented here). Once profiles are computed we can estimate both $r_{0,\text{scint}}$ and $r_{0,\text{sector}}$, and use the ratio $r_{0,\text{scint}}/r_{0,\text{sector}}$ for a \textit{prima facie} indication of accuracy. The radial motion method, $r_{0,\text{sector}}$, is partially independent from the measured optical turbulence profile used for estimating $r_{0,\text{scint}}$. This comparison is similar to the benefit offered from the common implementation of a combined MASS-DIMM instrument \cite{kornilov2007combined}. The norm of the residuals from the non-negative least squares solution to \autoref{eq:main_eq}, $\sigma_{NNLS}$, can also be used to indicate low quality data. We also compute profiles from approximately 2 minute rolling averages of the APS, which helps to constrain the solution and minimise zero estimates of $J_i$ \cite{tokovinin2007accurate}.

\begin{figure}[h]
  \centering
  \includegraphics[width=\linewidth]{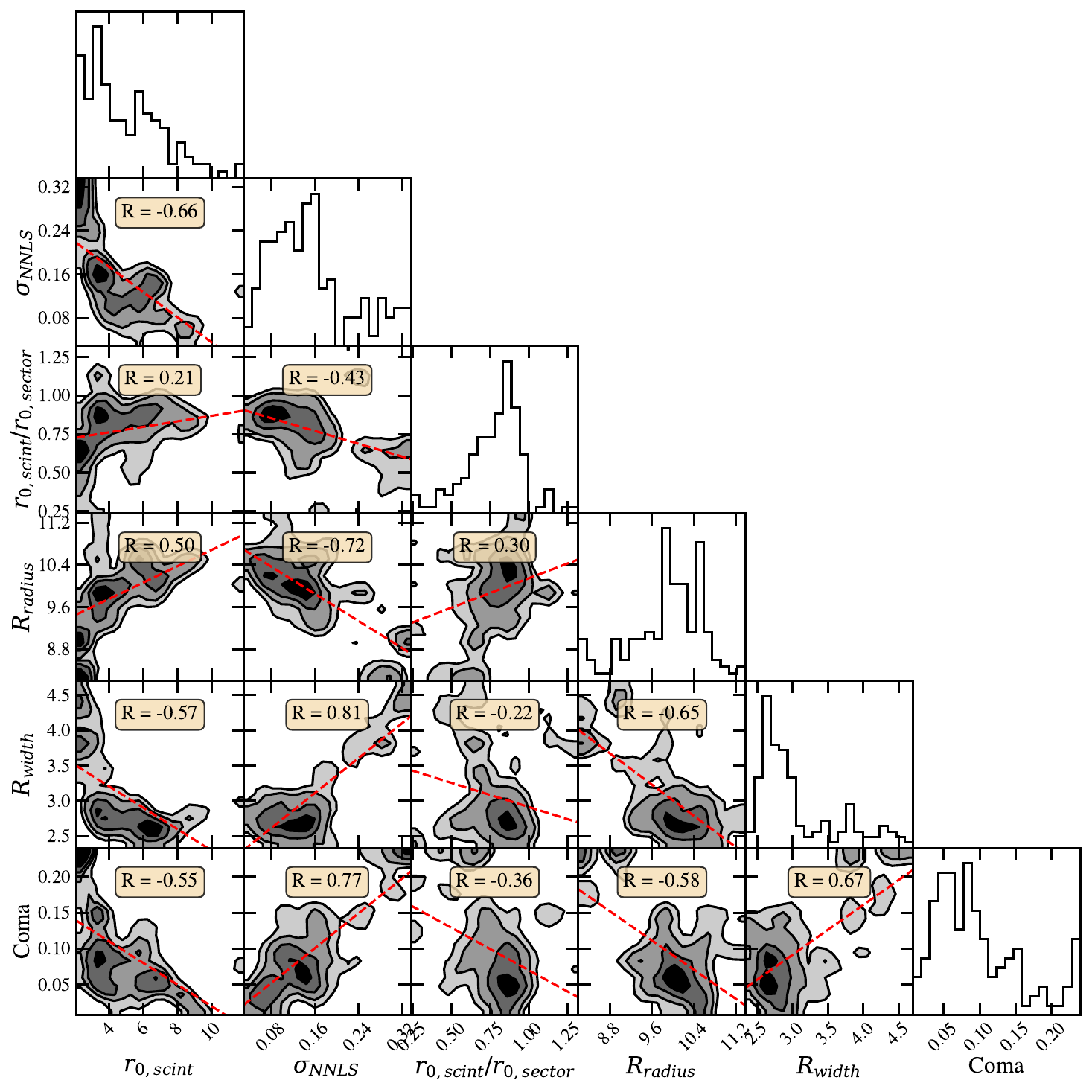}
  \caption{Corner plot displaying the correlation between a number of important parameters for data and image quality, measured from preliminary observations at TMF. Diagonal entries show the histogram of values corresponding to each columns parameter. Dashed red line is the line of best fit from a linear regression between parameters and the $R$ values are the Pearson correlation coefficients between parameters. $r_0$ is in centimetres, scaled to $500$~nm and zenith, as is done throughout this study.}
  \label{fig:quality_corner}
\end{figure}

Detecting low-quality data can be difficult if image quality typically degrades in strong turbulence, requiring careful calibration for any relations that might exist. We produce a corner plot, \autoref{fig:quality_corner}, to show correlations between parameters of interest. \autoref{fig:quality_corner} shows that $r_{0,\text{scint}}/r_{0,\text{sector}}$ is largely independent of turbulence strength but a loose correlation exists for other parameters with increasing turbulence (decreasing $r_0$). $\sigma_{NNLS}$ is correlated with image specification parameters in \autoref{fig:quality_corner}, and outliers of $\sigma_{NNLS}$ from the weak relation with $r_0$ were often found to correspond with poor image quality, e.g. weak signal from thin cloud or wind shake causing the spot to temporarily leave the region-of-interest. We also observe $R_\text{width}$ increases in stronger turbulence, possibly due to the blurring of rapid changes in $R_\text{radius}$ within exposures, but may also be caused by high wind speed. We aim to find more robust correlations between all parameters shown in \autoref{fig:quality_corner} to optimise performance and maintain data integrity.

\subsection{Dual $r_0$ comparison}
\label{sec:dualr0}

\begin{figure}[h]
  \centering
  \includegraphics[width=0.8\linewidth]{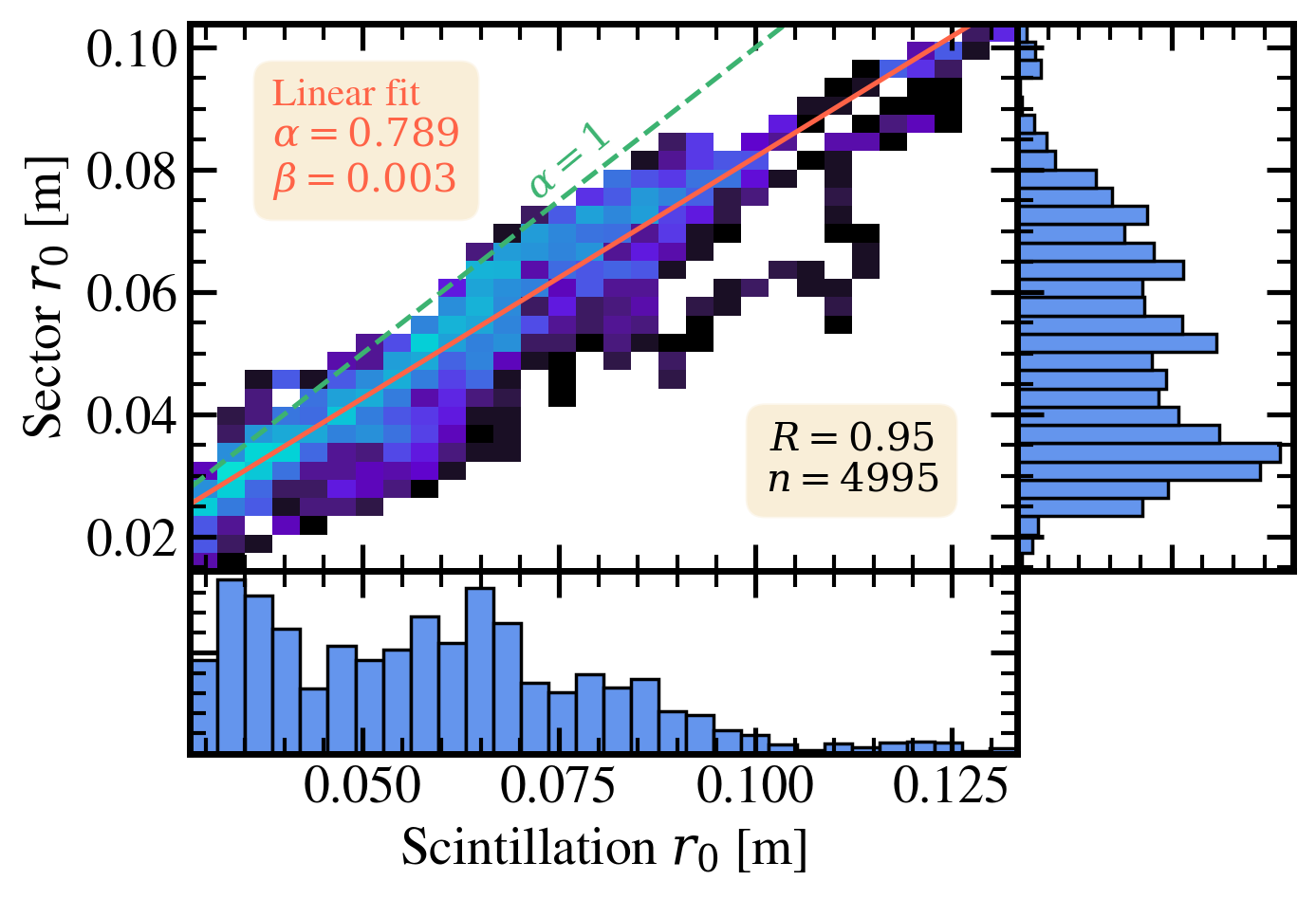}
  \caption{Comparison of alternate Fried parameter, $r_0$, measurement techniques with preliminary RINGSS measurements at TMF. Scintillation $r_0$ is the integral of turbulence profile. Sector $r_0$ derives from the radial motion method of $r_0$ estimation. Linear regression fit is shown in solid red line and compared with a slope ($\alpha$) of $1$ in dashed green. Pearson correlation coefficient, $R$, between measures is shown as $0.95$. Systematic offset, $\beta$, from linear regression is in units of metres.}
  \label{fig:sector_scint_ratios}
\end{figure}

\autoref{fig:sector_scint_ratios} highlights the relation between $r_{0,\text{scint}}$ and $r_{0,\text{sector}}$ over a range of turbulence strengths with measurements conducted in daytime and nighttime at TMF. Although \autoref{fig:sector_scint_ratios} displays a robust relation between these measurement techniques, there is a significant trend of $r_{0,\text{sector}}<r_{0,\text{scint}}$, i.e. overestimating turbulence strength. $r_{0,\text{sector}}$ is typically expected to be larger than $r_{0,\text{scint}}$, as described in \cite{tokovinin2021measurement}, and the opposite result here could be caused by numerous issues we discuss here. Radial sector motion, like with a DIMM, can overestimate free atmosphere turbulence, an effect enhanced by astigmatism that is described in \cite{tokovinin2007accurate}. However, error in the expected pixel scale is more likely as it cannot be easily measured in this system due to the narrow linewidth filter blocking all stars, i.e. plate-solving or binaries cannot be observed to measure pixel scale. The larger than expected $R_\text{width}$ we observe also indicates an out of specification focal length and therefore pixel scale.

Another possible cause is the profile restoration underestimating turbulence due to scintillation saturation, so more data in weaker turbulence could be used to assess this theory \cite{tokovinin2021measurement,tokovinin2007accurate}.  Nonetheless, further observations are required to draw conclusions, though we consider the strong correlation ($R=0.95$) between methods a good first indicator the instrument is producing reliable results.

\section{Accuracy analysis}
\label{sec:accuracy}

\begin{figure}[h]
  \centering
  \includegraphics[width=0.7\linewidth]{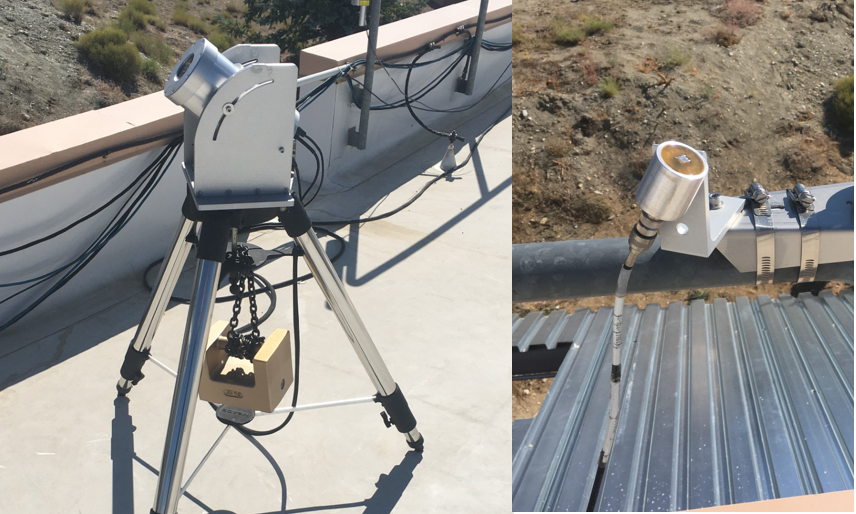}
  \caption{Integrated turbulence monitors at Table Mountain Facility. Left: Polaris motion monitor \cite{C.cavadore}. Right: solar scintillometer \cite{beckers2001seeing}. Note that the Polaris motion monitor points North, while RINGSS and solar scintillometer point South.}
  \label{fig:alternate_instruments}
\end{figure}

We compare against three turbulence monitors at TMF to further assess RINGSS. Comparing turbulence monitors is often difficult due to different measurement techniques, positions on the mountain, and from the stochastic nature of turbulence, nonetheless this remains the best tool available \cite{tokovinin2023elusive}. TMF hosts a solar scintillometer, Polaris motion monitor, and boundary-layer scintillometer, the latter being covered separately in \autoref{sec:BLS_comp} \cite{van2015long,beckers2001seeing,C.cavadore}. The Polaris monitor is a single aperture instrument which derives $r_0$ from absolute image motion of Polaris captured at 50~Hz \cite{C.cavadore}. The solar scintillometer was previously calibrated against a solar DIMM and is mounted alongside the Polaris monitor on the roof of OCTL, approximately 50~m from RINGSS (see \autoref{fig:ringss_image}) \cite{beckers2001seeing}. All instruments, including RINGSS, are corrected to represent values at zenith and 500~nm for appropriate comparison. These integrated seeing monitors, located on the roof of OCTL, are shown in \autoref{fig:alternate_instruments}.

\begin{figure}[h]
  \centering
  \includegraphics[width=0.9\linewidth]{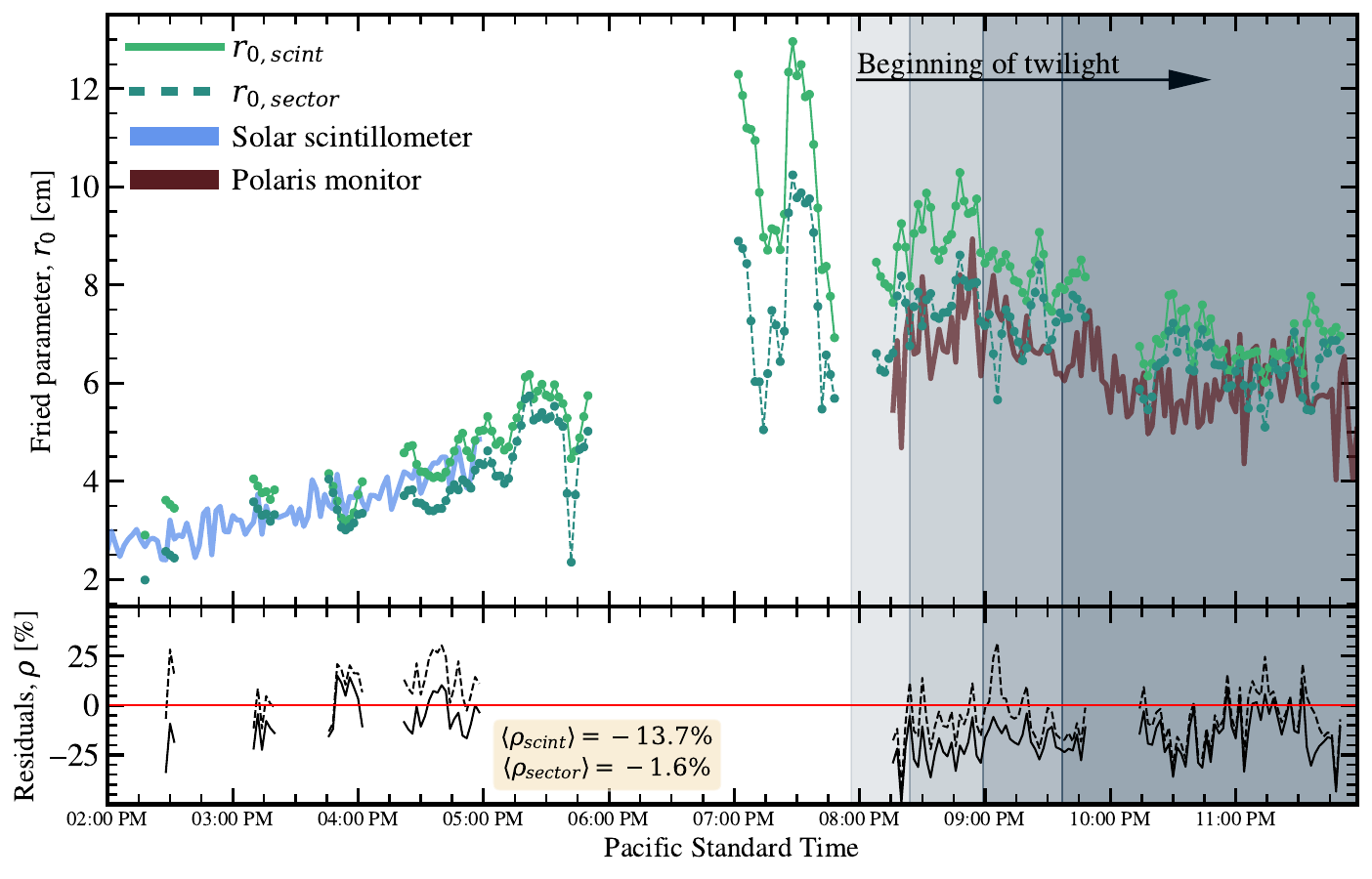}
  \caption{Sample of Fried parameter RINGSS measurements from the 24$^\text{th}$ of May 2023 demonstrating a strong correlation with two separate instruments over daytime and nighttime. Gaps in data are caused by a combination of technical issues and pauses in LCRD transmission. Lower inset shows the residuals between the solar scintillation and the Polaris monitor to RINGSS normalised against measured RINGSS $r_0$ values. The three phases of twilight and nighttime are shown in shaded regions toward the right.}
  \label{fig:r0_time_series}
\end{figure}

A nearly continuous observing block from the 24th of May 2023 is plotted in \autoref{fig:r0_time_series} to highlight the comparison with the solar scintillometer and Polaris monitor, in addition to showing the diurnal evolution of turbulence. Residuals are computed by time-averaging all methods into two-minute bins. Gaps of RINGSS data in \autoref{fig:r0_time_series} are due to technical issues and pauses in LCRD operations. There is also a break between when the solar scintillometer operates and the Polaris monitor is able to begin measurements. \autoref{fig:r0_time_series} shows that two different $r_0$ estimates from RINGSS agree with two separate instruments over the presented time range, including in strong turbulence. The daytime performance we see is encouraging as scintillation saturation is a major challenge for profilers which rely on weak-scintillation assumptions. However, scintillation is significantly weaker at $1550$~nm than typical observations in the visible, which is a major factor for this instruments reliability (see \autoref{sec:prof_results}). We nonetheless still apply a saturation correction as described in \cite{tokovinin2021measurement} and developed in \cite{tokovinin2007accurate}, though it's impact on $r_0$ estimation is $<5\%$.

\begin{figure}[h]
  \centering
  \includegraphics[width=0.6\linewidth]{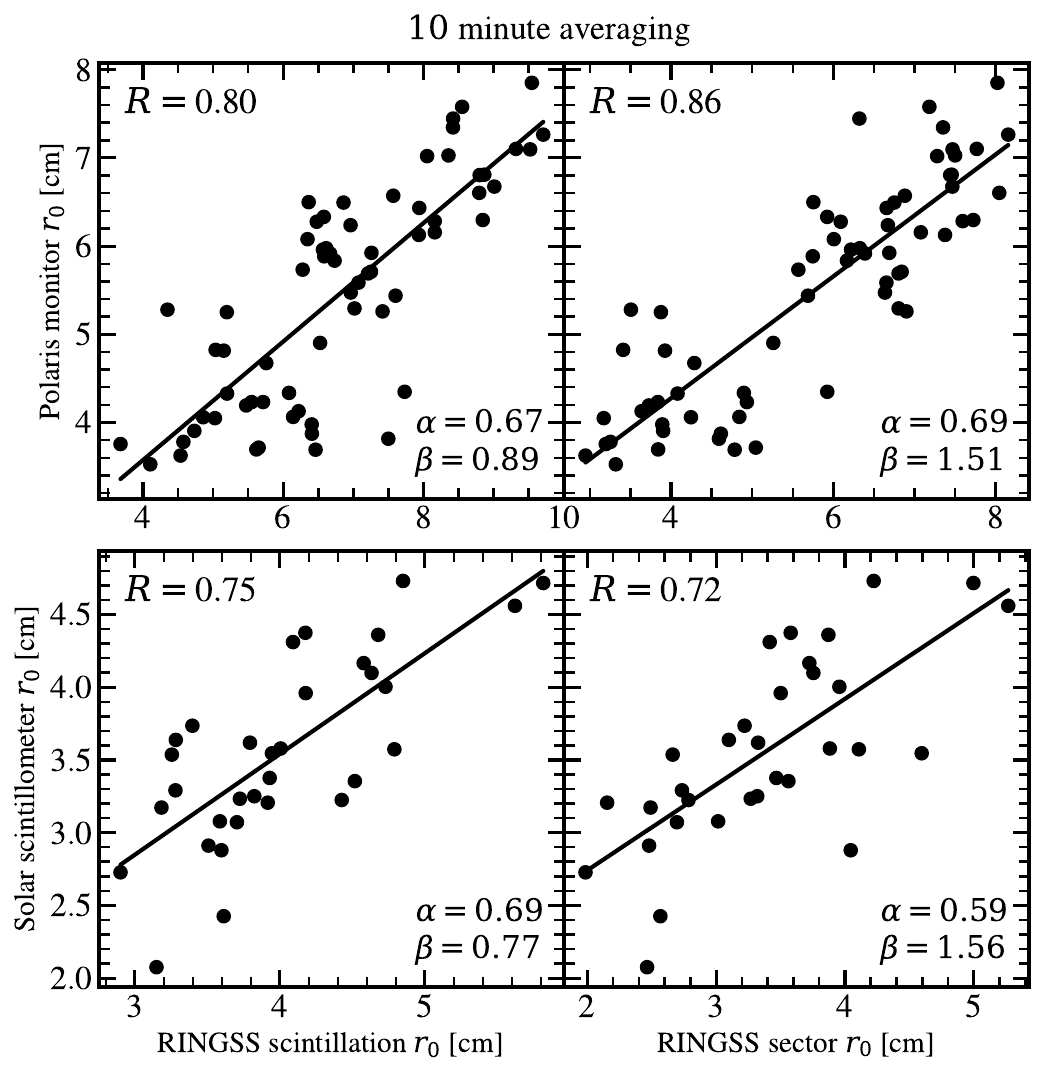}
  \caption{Correlation plots between both the RINGSS $r_{0,\text{scint}}$ and $r_{0,\text{sector}}$ with the Polaris monitor and solar scintillometer. Linear regression fits shown in solid black line. Annotated $R$ values represent the Pearson correlation coefficient. $\alpha$ is the fitted slope and $\beta$ (shown in units of cm) are the systematic offsets from the fit, i.e. the intercept.}
  \label{fig:correlation_subplots}
\end{figure}

We then compare between all instruments over the entire dataset, time-averaging to 10 minute bins for comparison as shown in \autoref{fig:correlation_subplots}. Although any degree of time-averaging is necessary for comparison, 10~minute bins are chosen partly because these instruments are not in the same position on the mountain, and not observing through the same path, hence we don't expect short term agreement ($\lesssim1$~minute). Further, we choose 10~minutes as an intuitive yet arbitrary scale to characterise current conditions by given the stochastic nature of turbulence. \autoref{fig:correlation_subplots} shows the Pearson correlation coefficients ($R$) and linear regression fits between RINGSS, the Polaris monitor, and solar scintillometer over this preliminary dataset. We observe reasonable correlation coefficients, i.e $R>0.7$ for all cases, promising in the context of turbulence monitors, but large systematic offsets ($\beta$) and slopes all significantly less than $1$. Nighttime RINGSS observations tends to agree more with the Polaris monitor than daytime RINGSS with the solar scintillometer. This could be either due to RINGSS being more accurate in weaker turbulence, e.g. due to insensitivity or increased error towards a strong boundary layer, or that the Polaris monitor is more accurate than the solar scintillometer. In all cases we observe RINGSS estimating stronger turbulence than alternate instruments, though more data is required to assess whether this is systematic.  Long term operation of RINGSS at TMF is required to see if these early promising correlations improve and hold over a broader range of conditions. Inherently different measurement techniques, instrument error, and different positions on TMF are all reasons to not expect a robust agreement between instruments in the long-term that might be expected of other meteorological parameters.

\section{Boundary layer measurements}
\label{sec:BLS_comp}

Effects of turbulence are typically worst in the daytime, where a convective boundary layer can form and dominate turbulence, so sensitivity to this layer is critical \cite{kaimal1976turbulence}. We are therefore interested in determining the accuracy of RINGSS near the surface in daytime, i.e. limiting turbulence conditions in daytime operation. To assess RINGSS accuracy in this regime we compare with a SCINTEC BLS900 boundary layer scintillometer (BLS), which is installed on the roof of OCTL (nearby to instruments shown in \autoref{fig:alternate_instruments}) to sample $C_n^2$ at high frequency in a horizontal path at all times of day \cite{van2015long}. The BLS points south and measures turbulence over a valley to a nearby mountain $3.4$~km from TMF in the same direction that RINGSS observes LCRD, though its location at OCTL is raised off the ground unlike RINGSS. We compare BLS measurements with RINGSS turbulence boundary layer estimates by dividing the lowest $J_i$ integral by layer height, i.e. from $0$~m to $250$~m, to estimate $C_n^2$. This method assumes a uniform distribution of $C_n^2$ through the layer but given RINGSS is observing over a steep valley and at a slant path we cannot justify a more sophisticated solution.

\begin{figure}[h]
  \centering
  \includegraphics[width=0.7\linewidth]{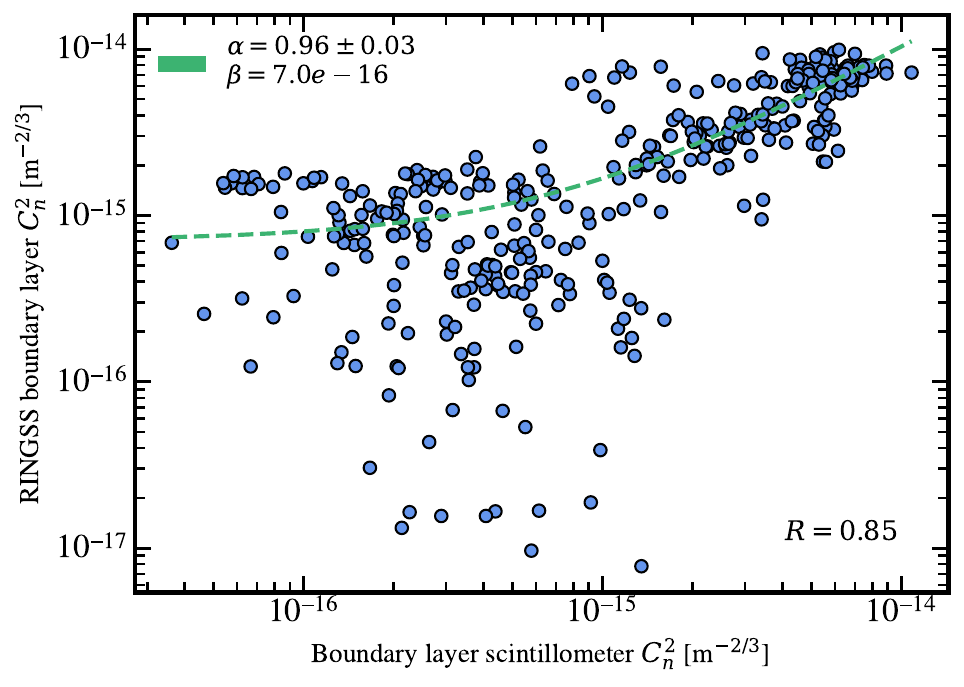}
  \caption{Comparison of simultaneous measurements of boundary layer turbulence from a boundary layer scintillometer and RINGSS at TMF during daytime and nighttime. Pearson correlation coefficient, $R$, is displayed on lower right. Linear regression is shown with green dashed line with slope, $\alpha$, and intercept, $\beta$. RINGSS boundary layer estimate is the average $C_n^2$ from $0$~m to $250$~m obtained by dividing the first profile layer by layer thickness.}
  \label{fig:boundary_correlations}
\end{figure}

\autoref{fig:boundary_correlations} shows the distribution of $C_n^2$ from the BLS compared with the ground layer $C_n^2$ estimate from RINGSS. We observe agreement within expectations for differing instruments measuring $C_n^2$, though with heteroscedasticity towards small $C_n^2$ likely due to a lack of data and different paths through the boundary layer. Nonetheless, \autoref{fig:boundary_correlations} demonstrates that RINGSS is capable of approximately estimating the boundary layer turbulence even in daytime. However, it should be noted that this preliminary dataset represents fairly weak boundary layer turbulence, even at daytime, likely due to the steep valley, altitude of TMF, and cool weather. Significantly more data is required over a substantial dynamic range to established a robust correlation with RINGSS and the BLS, particularly for $C_n^2>10^{-15}$~m$^{-2/3}$. Correlation of boundary layer turbulence with wind-speed and direction from meteorological instruments will also be considered in future work.

\section{Profile results}
\label{sec:prof_results}

\begin{figure}[h]
  \centering
  \includegraphics[width=\linewidth]{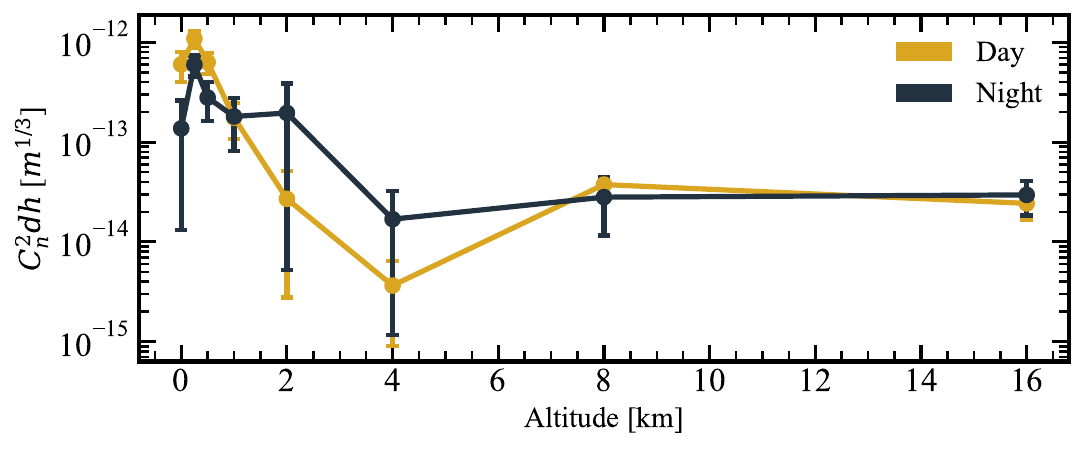}
  \caption{Median turbulence profiles measured by RINGSS at the Table Mountain Facility, California. Day and night time profiles are split to highlight the diurnal evolution of turbulence, particularly in the boundary layer. Error bars are calculated from the median absolute deviation.}
  \label{fig:diurnal_profs}
\end{figure}

We show median turbulence profiles measured by RINGSS over the preliminary dataset in \autoref{fig:diurnal_profs}, split into daytime and nighttime. The median is at each layer among non-zero values to ignore missing layers caused by non-negative least squares. Turbulence profiles are the series of $J_i$ integrals representing $J_i=\int_{z_i}^{z_{i+1}} C_n^2(h)dh$ with units $m^{1/3}$. \autoref{fig:diurnal_profs} shows the characteristic dissipation of boundary layer turbulence at night time, and a moderate increase in the free atmosphere. Weak daytime turbulence in the free atmosphere, shown in \autoref{fig:diurnal_profs}, could be caused by strong turbulence in the boundary layer, a blocking effect which has been investigated with MASS. Note that we do not sample the full range of daytime and nighttime due to the constraints of LCRD operations, daytime is typically morning and nighttime typically early evening. As RINGSS continues to operate and develop into an automated system at TMF, we aim to build robust statistical estimates of the profile. Profiling also provides the ideal input for any turbulence forecasting systems, to be explored in future work, as it can be paired with meteorological forecasts of wind speed and temperature profiles \cite{osborn2018atmospheric}. Although we have presented comparisons of RINGSS against integrated turbulence monitors, a critical part of future work will be to compare profiles such as in \autoref{fig:diurnal_profs} against other daytime turbulence profilers, e.g. the 24hSHIMM \cite{griffiths2023demonstrating}.

Turbulence profiling also allows us to estimate a number of important integrated parameters, particularly important for AO operation at OCTL during LCRD links. \autoref{tab:params} shows the median atmospheric coherence time ($\tau_0$), Greenwood frequency ($f_G$) and isoplanatic angle ($\theta_0$) from all measured profiles in the preliminary dataset, split into daytime and nighttime. Estimating $\tau_0$ and $f_G$ requires knowledge of the wind speed profile, though the turbulence-weighted wind velocity, $\bar{V}$, can be substituted for some loss of accuracy \cite{kellerer2007atmospheric}. \autoref{eq:tau_0} shows how $\tau_0$ and $f_G$ are computed for $\bar{V}$ if wind speed profiling, e.g. measured by a Stereo-SCIDAR, is not available \cite{osborn2016turbulence}. We also include an estimation of the Rytov variance, $s_{0}^2$, as formulated in \cite{roddier1981v} for the small-signal approximation (RINGSS cannot directly measure the scintillation index like the solar scintillometer). $s_{0}^2$ is estimated for the operating wavelength of $1550$~nm and highlights the benefit of observing in infrared for remaining in the weak to moderate scintillation regime, i.e. $\sigma_{0}^2\approx0.06$ as in \autoref{tab:params} would be $\approx0.2$ if observed at $550$~nm. It was shown in \cite{tokovinin2021measurement} that depending on the profile structure, even a moderate $s_{0}^2$ can be problematic for resolving layers close to the surface.

\begin{equation}
    \tau_0=0.314\frac{r_{0,\text{scint}}}{\bar{V}} \: \: \: \: \: f_G = 0.43\frac{\bar{V}}{r_{0,\text{scint}}}
    \label{eq:tau_0}
\end{equation}

To estimate $\bar{V}$ with RINGSS we instead use the atmospheric second moment of the wind, $V_2$, described by Kornilov in \cite{kornilov2011stellar} with implementation for RINGSS described in \cite{tokovinin2021measurement}. $V_2$ is principally measured with RINGSS from relating fast focus variation to the temporal structure function for atmospheric piston-modes, a concept with origins in the FADE instrument, and has been shown to follow $V_{5/3}$ closely, which defines $\tau_0$ \cite{tokovinin2008fade}. We do not reproduce this methodology here, as it is involved and best explained by Kornilov \cite{kellerer2007atmospheric,tokovinin2008fade,kornilov2011stellar}. Measuring $V_2$ from rapid changes in defocus caused by scintillation as $R_\text{radius}$ is sensitive to blurring so the short exposure time, $200$~\textmu s, afforded by the strong LCRD downlink, is significant in improving accuracy here.

\begin{table}[h]
    \centering
    \caption{Median integrated parameters from RINGSS turbulence profile measurements at TMF. Data from short preliminary dataset is split into daytime and nighttime LCRD links. Uncertainties are the standard deviation from all parameter calculations over the dataset}
    \begin{threeparttable}
        \begin{tabular}{c|c|c|c|c|c|c} 
             & $r_{0,\text{scint}}$ [cm] & $r_{0,\text{sector}}$ [cm]& $f_G$ [Hz]& $\theta_0$ [arcsec]& $\tau_0$ [ms] & $\sigma_0^2$$\:^{\dagger}$\\ 
            \hline
            Night & $7.2\pm1.8$ & $6.0\pm1.6$ & $55\pm19$ & $1.7\pm0.5$" & $2.7\pm0.8$ & $0.063\pm0.02$\\
            \hline  
            Day & $5.0\pm1.4$ & $4.3\pm1.2$ & $68\pm33$ & $2.1\pm0.4$ & $2.2\pm0.4$ & $0.064\pm0.03$\\
        \end{tabular}
        \begin{tablenotes}
            \item[\textdagger] Rytov variance, $\sigma_0^2$, is computed for the measurement wavelength of $1550$~nm, rather than $500$~nm like other integrated parameters \cite{roddier1981v}.
        \end{tablenotes}
    \end{threeparttable}
    \label{tab:params}
\end{table}

In \autoref{tab:params} we see almost all parameters degrade in daytime compared with nighttime, however $\theta_0$ worsens slightly at nighttime due to the increased turbulence above 2~km visible in \autoref{fig:diurnal_profs}. This highlights one of the benefits of profiling over integrated turbulence measurements, i.e. $\theta_0$ is an important parameter to consider for uplink pre-compensation with AO which is driven by high altitude turbulence \cite{martinez2023atmospheric}. The Rytov variance approximation from \cite{roddier1981v} is also less sensitive to increases of turbulence strength near the surface which explains the very similar $\sigma_0^2$ values between day and night.

\section{Conclusion}

We have successfully demonstrated the deployment of a RINGSS turbulence profiler, an instrument initially demonstrated in 2021, and conducted the first turbulence profiling with a laser communications satellite using the LCRD GEO optical downlink signal \cite{tokovinin2021measurement}. This is also the first time that a RINGSS instrument has been used in daytime and strong turbulence conditions, highlighting the measurement technique is robust over a broad range of conditions. Preliminary results have been compared with a suite of other turbulence monitors at TMF, showing reasonable correlations ($R=0.75-0.86$) despite the small dataset and difficulties of finding agreement among turbulence instruments. Efforts are also ongoing to automate this instrument with closed-loop tracking, auto-focusing, and an observatory control system.  While we currently lack the data to make broad and robust statements on the instrument performance and site characteristics we have demonstrated the unique concept of FSOC turbulence profiling which could improve the performance of GEO feeder networks. Future operations of RINGSS at TMF alongside LCRD operations will aim to constrain the correlations we have presented here and develop this instrument into a capability for improving reliability with short-term turbulence prediction and inform the AO control system on OCTL. 

\begin{backmatter}
\bmsection{Funding}
Funding for MB to undertake this research was provided by the Australian Commonwealth Scientific and Industrial Research Organisation (CSIRO) under the postgraduate supplemental scholarship scheme. MB is also supported by the Australian Research Training Program.

\bmsection{Acknowledgments}
The research was carried out at the Jet Propulsion Laboratory, California Institute of Technology, under a contract with the National Aeronautics and Space Administration. \\ The authors would like to thank the LCRD and OCTL operational teams. We also thank Andrei Tokovinin at the National Optical-Infrared Astronomy Research Laboratory for providing advice and input toward implementing the RINGSS instrument.\\

Portions of this work were presented at the SPIE Photonics West Free-Space Laser Communications XXXVI conference in 2024 as \textit{``Measuring the vertical profile of atmospheric turbulence with the laser communication relay demonstration downlink at Table Mountain Facility}'' \cite{birch2024measuring}.

\bmsection{Disclosures}
The authors declare no conflicts of interest.

\end{backmatter}

\bibliography{sample}

\end{document}